\documentstyle[12pt]{article} \headheight=12pt \footskip=48pt
\jot=8pt \newcommand{\an}{\&} \newcommand{\be}{\begin{equation}}
\newcommand{\ee}{\end{equation}} 
\newcommand{\eei}{\end{equation}\indent\indent}
\newcommand{\bc}{\begin{center}} \newcommand{\ec}{\end{center}}
\newcommand{\ber}{\begin{eqnarray}} \newcommand{\ear}{\end{eqnarray}}
\newcommand{\ba}{\begin{array}} \newcommand{\ea}{\end{array}}
\newcommand{\n}{{}^{(3)}\nabla} \newcommand{\na}{\nabla}
 \newcommand{\hs}{\,-\,}

 \newcommand{\su}{{\cal U}}

\newcommand{\sfrac}[2]{{\textstyle{#1\over#2}}}
\def\case#1/#2{\textstyle\frac{#1}{#2} }
\begin{document}
\begin{titlepage}
  \hoffset=-37pt \title{Dust\hs Radiation Universes: Stability
    Analysis}

\author{{\sc Marco Bruni}\thanks{Present
address: School of Mathematical Sciences,
Queen Mary and Westfield College, Mile End Road E1 4NS, London, U.K.}
 \\ 
\normalsize{\it School of Mathematical Sciences,
Queen Mary and Westfield College, London U.K.}
\\
\normalsize{\it Dipartimento di Astronomia, Universit\`a di Trieste,
Italy.}
         \\  \\ 
{\sc Kamilla Piotrkowska}          \\ 
\normalsize{\it S.I.S.S.A. - I.S.A.S. via Beirut 2/4 Trieste, Italy}
\\ \\ 
                      } 
\date{$\mbox{}$ \vspace*{0.3truecm} \\ \normalsize{\today}           }
\maketitle
\thispagestyle{empty}
\vspace*{0.5truecm}
\begin{abstract}

  In this paper we consider flat and open universe models containing a
  mixture of cold matter (dust) and radiation interacting only through
  gravity, with the aim of studying their stability with respect to
  linear scalar perturbations. To this end we consider the perturbed
  universe as a dynamical system, described by coupled differential
  equations for a gauge\hs invariant perturbation variable and a
  relevant background variable.

The phase\hs space analysis of this dynamical system shows
that flat dust\hs ra\-dia\-tion models are unstable,
and open models structurally unstable, with respect to adiabatic
perturbations. For flat models, there are actually three different
regimes of evolution for the perturbations, depending on their
wavelength
and the transition
scale from one  regime to the other is determined by a critical wavenumber
for the perturbations, $k_{EC}$ (an invariant of the model).
We find that  $k_{EC} < k_{JE}$ (the Jeans wavenumber at equidensity
of matter and radiation), implying that there are perturbations
which decay even if their wavelength at equidensity is larger than
the corresponding Jeans scale.
We also briefly discuss metric and curvature perturbations.

We believe that this analysis gives a clearer idea of the stability
properties of realistic universe models than
the standard one based on the Jeans scale, despite our
simplifying assumptions.
%
\vspace*{0.2truecm}
\bc Ref. SISSA 126/93/A      
 (July 30,1993)\ec              
 Feb. 90
\end{abstract}
\end{titlepage}
%
%

 \section{Introduction}

 A main goal of present\hs day cosmology is to understand the
 formation of the structures (galaxies, clusters, superclusters)
 observed in the universe, while trying to explain why, on large
 enough scales, this seems to be so well described by the
 Friedmann\,-\, Lema\^\i tre\,-\, Robertson\,-\, Walker (FLRW
 hereafter) isotropic models. Given these latter, most theories of
 structure formation are based on the gravitational instability
 scenario. At any given epoch, there are perturbations larger than a
 certain characteristic - time dependent - scale\footnote{This can be
   the Jeans scale, or the Hubble radius sometimes loosely referred to
   as horizon; a more important scale is actually the sound horizon,
   see \cite{bi:bardeen}.}; while perturbations much smaller than this
 scale oscillate as sound waves, the larger density perturbations
 grow, eventually entering a non\hs linear regime during the matter
 dominated epoch, thus forming the observed structures.  The
 mathematical basis for such scenario is the theory of perturbations
 of FLRW models.

In the inflationary scenario, the
perturbations were generated from quantum fluctuations within the
Hubble horizon $H^{-1}$, have  evolved classically outside the horizon, and
have re-entered it during the radiation or  matter dominated epochs
(perturbations with larger wavelength re-entering later).
During the last decade, observations of the distribution of matter
have shown that the scale at which the background homogeneity is
reached is larger than what was thought before, being of the order of
hundreds of Mpc\footnote{The answer to the question
``At which scale in the universe there is a
transition to homogeneity?" depends on how the question is posed, i.e.
how such a scale  is defined. In a very loose
sense we can say that this scale is measured by the size of the
largest structures we see (of the order of $10^2$ Mpc), and in a strict
sense such a scale does not exist at all if there is not a cut-off in
the perturbation spectrum. Various reasonable definitions of such a
scale can be given (e.g. specifying a level for the
fluctuations, such that we can say that there is a transition to
homogeneity at the scale where the fluctuations fall below the
specified threshold): see for examples  \cite{bi:rob1}
 and \cite{bi:einasto}.}.
 Consequently, both the inflationary scenario and the
large scale observations motivate the use of a fully relativistic
theory of perturbations in FLRW models
in order to study the formations of the
larger structures.\footnote{Newtonian perturbation theory being
applicable only for vanishing pressure and at scales much smaller than
the horizon.}

For the latter ones, the density contrast against the homogeneous
background appears to be small enough that a linear perturbation
analysis still suffice to describe the evolution of perturbations of
the corresponding scale. Even if the density contrast is mildly non\hs
linear, the curvature perturbations are still in the linear
regime \cite{bi:barrow}, \cite{bi:fagundes}, thus
one can imagine the present universe as well described by a linearly
perturbed FLRW model at large scales, while non\hs linearities at
smaller scales can be considered smoothed out in this
picture.\footnote{In general, this raises the issue of defining a proper
averaging procedure, here we do not address this problem; see e.g.
\cite{bi:fut} or \cite{bi:jacobs}.}
Then the question arises if  in this respect  we live in a special epoch - an
epoch in which large scale perturbations are still in the linear regime
- or if this is a natural output in the theoretical context of
perturbed FLRW models. This question is of the same sort as that posed
by the "flatness problem": if the geometry of the universe is non-flat,
then we live in a special epoch in which the density parameter is still
close to unity $\Omega_{o} \sim 1$.
It is well known that in a flat dust model
there is a  perturbation mode that  grows unbounded, while in an open
dust universe there is a mode that  freezes in at an epoch $z\approx
\Omega^{-1}$. In both cases the perturbation equation (\ref{eq:1})
is the same for all modes, because for $c_{s}^{2}=w=0$ (dust) the
coefficient $\beta$ (\ref{eq:beta}) does not depend on the
wavenumber $k$. It is usually said that in a dust universe each
perturbation evolves as a separate FLRW universe. On the other hand,
perturbation scales in a pure radiation model always come within the
"sound horizon" (see \cite{bi:bardeen}) and oscillate
as sound waves.

This simple picture gives however rather little information about the
generality of this behaviour for the perturbations, and on the
stability properties of the perturbed FLRW model.

The aim of this paper is therefore to study the problem of the stability of
FLRW
models following an alternative approach: instead of looking for
analytic solutions (either exact or approximate, see e.g.
\cite{bi:padma}) of the perturbation equations, we consider the
perturbed universe as a dynamical system, described by coupled
differential equations for a gauge\hs invariant perturbation variable
and a relevant background variable. In this approach, the evolution of
perturbations is represented by the trajectories in the phase space of
the dynamical system, and  their final fate
is linked to the presence and the nature of the
critical points of the system.

In a certain sense the present work extends that of \cite{bi:ER} to
include perturbations, although we restrict our analysis to open and
flat models only, and we take a vanishing cosmological constant
$\Lambda=0$.
Previous works have followed the approach to stability used here
\cite{bi:wos}, \cite{bi:bruni}, but they only considered either a
pure radiation  or a pure  dust model. However, since the analysis
here is based on the study of the dynamical system at time\hs
infinity, the pure radiation model does not appear  physically
significant in this limit, while the stability properties of the dust
models are affected by the simplifying assumption of the vanishing of
the speed of sound in these models. Therefore here  we consider a
class of perturbed models containing both dust and radiation
as a more realistic
description of the real universe.
The dust component can be taken to represent Cold Dark Matter (massive
weakly interacting particles), while the radiation component represents
photons and other massless particles such as massless neutrinos.
Therefore these two fluids interact only through gravity.

Since our aim is to study the
stability properties of these simple  models, we consider only the total
density
perturbation (the single component perturbations are not directly
relevant to the evolution of curvature perturbations), assuming
adiabatic perturbations.

For the case of flat universe models we find that
 there exists a critical wavenumber
$k_{EC}$, which  is an invariant characteristic of the model
and is related to the only scale entering the flat models, i.e.
the Hubble radius at equidensity of matter and radiation $H^{-1}_E$.
The corresponding critical scale $\lambda_{EC}$
($\lambda_{E}$ is the perturbation wavelength at equidensity),
remarks the transition  from stability to instability,
 but in a way which is more rigorous - from the
point of view of the stability analysis - than the Jeans
or the Hubble scale.
We find that there are
actually three regimes for the evolution of fluctuations:
\begin{enumerate}
\item growing large scale perturbations (unstable modes)
\item overdamped intermediate scale perturbations
\item damped small scale wave perturbations,
\end{enumerate} where the transition scale from one regime to another
is always of the order of $\lambda_{EC}$, and for this latter,
separating cases $1.$ and $2.$, we find that the perturbed FLRW models
are {\it structurally unstable} \cite{bi:AP}.  Also, we show that
$\lambda_{EC}$ is of the order of the Jeans scale $\lambda_{JE}$ at
equidensity in the same model: however since $\lambda_{EC} \simeq 2.2
\lambda_{JE}$ (today $\lambda_{0C} \simeq 67 Mpc$), our analysis shows
that there are perturbation modes that decay, despite that their scale
$\lambda_{E}$ is larger than $\lambda_{JE}$ at equidensity.  Thus the
evolution of perturbations in these models depends on their scale, in
such a way that smaller scales evolve like in a pure radiation model
(case $3.$) and larger scales like in a pure dust model (case $1.$),
while we found a small intermediate range of scales (case $2.$) for
which perturbations are overdamped (critical damping occurs for the
transition scale between cases $2.$ and $3.$), which is an original
feature of the dust--radiation models, and to our knowledge was not
known before.

For the case of an open dust--radiation model instead, the evolution
of perturbations appears to be dominated by the curvature of the
background, and their final state is similar to that they have in
a pure dust model, i.e. all the perturbation scales are frozen in
to a constant value.
These models are {\it structurally unstable} \cite{bi:AP} with
respect to perturbations of any wavelength.

The plan of the paper is as follows: in section \ref{sec:dyn} we give
a brief outline of the method used here to study the stability of FLRW
models; in section \ref{sec:dyndr} we briefly describe the dust -
radiation models and we specify the perturbation equations for them;
in section \ref{sec:res} we describe our results and we compare them
with the Jeans stability analysis, also we briefly discuss metric and
curvature perturbations.  Finally, in section \ref{sec:conc} we give a
summary and discuss other possible applications of the method used
here.

In the following we take $c=1$, and $\kappa=8\pi G$ ($G$ is the gravitational
constant), and we assume a vanishing cosmological constant $\Lambda=0$.

\section{The dynamical system}\label{sec:dyn}

In a recent series of papers, Woszczyna and colleagues considered the
dynamics of Newtonian \cite{bi:woskz} and relativistic \cite{bi:wos}
linearly perturbed universe models, studying the stability of these
dynamical systems.  The analysis of the relativistic case was however
affected by a wrong assumption on the allowed range for a scale
parameter $k$ ($q$ in \cite{bi:wos} and in \cite{bi:bruni}) and by a
misinterpretation of the perturbation variables, as it was shown in
\cite{bi:bruni}.

In this article, we shall focus on relativistic perturbations of flat
and open FLRW  models containing two uncouple fluids coupled only
through gravity: dust and
radiation. Open and flat FLRW models expand for an infinite amount of
time and therefore one can apply standard stability criteria: in particular
one can establish if a cosmic dynamical system, i.e. a perturbed
cosmological model, is able or not to return to the original
homogeneous state
(in the latter case we have an instability in the sense of Liapunov
and the system is said to be structurally unstable \cite{bi:AP})
or  if the
perturbations grow indefinitely (instability in the Laplace
sense).

 In general, in the gauge\,-\, invariant approach to cosmological
perturbations the scalar\footnote{
It is standard to call scalar perturbations those related with density
perturbations describing the clumping of matter
(see e.g. \cite{bi:stewart}). In this paper we
shall consider only these perturbations, as the only relevant to the
problem of stability  of the universe.} density  perturbation variable
(we shall consider specific perturbation measures later) satisfy a
second order (in some time variable) differential equation. If one
restricts the attention to the harmonic component $X$ of the
perturbation, and assumes that this is adiabatic (see section
\ref{sec:pert}), its evolution is
given by a homogeneous ordinary differential equation

\begin{equation}
  \ddot{X} +\alpha(t)\dot{X} +\beta(t)X=0\;,\label{eq:1}
\end{equation} where $t$ here is proper time, and the dot indicates a
derivative with respect to $t$. When it is not possible to find a
simple solution to (\ref{eq:1}), a qualitative analysis of its
properties is useful in order to determine the late - time behaviour
of the perturbations\footnote{One can obviously find numerical
  solutions, but the study of the phase space allows us to obtain
  general conclusions for a generic set of initial conditions.}. The
coefficients $\alpha$ and $\beta$ are functions given by the
background dynamics, but in general their time dependence cannot be
explicitly determined.  Therefore, it is useful to think of $\alpha$
and $\beta$ as known functions of one or more parameters, and add to
(\ref{eq:1}) the evolution equation for the parameters in order to
have an autonomous system. A sensible choice followed in \cite{bi:wos}
is given by the density parameter $\Omega$, which in a FLRW model
satisfies the equation

\begin{equation}
  \dot{\Omega}=\Omega(\Omega-1)(\case{1}/{3} +w)\Theta\;, \label{eq:2}
\end{equation} where $\Theta=3 H=3\dot{a}/a$ is the expansion of the
cosmic fluid, $a$ is the FLRW scale factor, $H$ is the Hubble
parameter, and $w=p/\mu$ is the ratio of the pressure to the energy
density. It is useful to change the independent variable from the
proper time $t$ to a function of the
scale factor $a$; with the choice of\footnote{
There are various typos in \cite{bi:wos}: there, the power appearing here in
the definition of $\tau$ is missed in equation (3), and consequently a
factor $1/3$ is missed in equation (9).}
  $\tau=\ln a^3$, ($\frac{d\tau}{dt}=\Theta$)  equations
(\ref{eq:1}) and (\ref{eq:2}) give

\begin{equation}
  X'' +\psi X' +\xi X=0\;,\label{eq:X}
\end{equation}

\begin{equation}
  \Omega'=\Omega(\Omega-1)(\case{1}/{3} +w)\;, \label{eq:3}
\end{equation} and in general

\be w'= -(1+w)\left[c_s^2(w) -w\right]\; \label{eq:w} \ee gives the
evolution of $w$, and in the previous equations the prime refers to
the derivative with respect to $\tau=\ln a^{3}$.

In a single fluid FLRW model the dynamics
is fixed by an equation of state $p=w\mu$, with $w=const$ (e.g. $w=0$
dust, $w=1/3$ radiation), and in this case   $c_s^2=\dot{p}/\dot{\mu}$
is the speed of sound, and $c_s^2=w$. In dealing with two or more
fluids however, $w'\not =0$, and $c_s^2\not = w$ is no longer the speed
of sound unless the fluids are coupled,
and $\psi$ and $\xi$ are given by
\begin{eqnarray}
  \psi(w,\Omega) &=&
  \frac{\dot{\Theta}}{\Theta^2}+\frac{1}{\Theta}\alpha(w,\Omega) =
  -\frac{1}{3} -\frac{1}{6}(1+3w)\Omega
  +\frac{1}{\Theta}\alpha(w,\Omega)\;,~~~~~\label{eq:4} \\
  \xi(w,\Omega) &=& \frac{1}{\Theta^2}\beta(w,\Omega)\;, \label{eq:5}
\end{eqnarray} while the second step in (\ref{eq:4}) is given by the
Raychaudhuri equation ($\kappa=8\pi G$) \be \dot{\Theta} +\sfrac13
\Theta^2 +\sfrac12\kappa\mu(1+3w) =0\;,
\label{bi:ray}
\ee governing the evolution of $\Theta$ (see e.g. \cite{bi:DBE}). In
the simplest case of one single fluid \cite{bi:wos}, \cite{bi:bruni}
with $w=const$ we have a third order autonomous system, given by
(\ref{eq:X}) and (\ref{eq:3}) (or a corresponding pair of first order
equations); also, equation (\ref{eq:3}) forms an autonomous first
order subsystem in this case. In the most general case we have a
fourth order autonomous system, with (\ref{eq:w}) as autonomous
subsystem (as we shall see in section \ref{sec:back}, $c_s^2=c_s^2(w)$
is fixed once the fluid components are specified). The order of the
system can however be reduced: first, it turns out that in the
practical case (see section \ref{sec:back}) either $w$ or $\Omega$ can
be eliminated; second, here we are not really interested in the
evolution-law of $X$, but rather in its qualitative behavior. Because
of this, we can achieve a further dimensional reduction of the phase
space passing to the Riccati equation corresponding to (\ref{eq:X}).
Introducing $Y=X'$ and

\begin{eqnarray}
  {\cal U} & = & Y/X=X' /X\;, \label{eq:sudef} \\ {\cal R} & = &
  \sqrt{X^2+Y^2}\;,
\end{eqnarray} we pass in the new phase space $\{{\cal R},{\cal
  U},w,\Omega\}$, where
\begin{equation}
  \frac{{\cal R}'}{{\cal R}}=\frac{{\cal U}(1-\xi-\psi{\cal U})}{{\cal
      U}^2+1}\;,
\end{equation}
\begin{equation}
  {\cal U}'=-{\cal U}^2-\psi{\cal U}-\xi\;, \label{eq:suev}
\end{equation} while the evolution of $\Omega$ and $w$ is still given
by (\ref{eq:3}) and (\ref{eq:w}). The variable ${\cal R}$ represent an
``amplitude" of the perturbation and it is not directly relevant to
the present analysis. Since (\ref{eq:suev}) and (\ref{eq:3}),
(\ref{eq:w}) form an autonomous subsystem, one can restrict the
analysis to the phase space $\{{\cal U},w,\Omega\}$; moreover, as we
said above, in practical cases we can restrict our attention either to
$\{\su,w\}$ or to $\{\su,\Omega\}$. The relevant variable here is
${\cal U}$: when it is positive we have either a growing density
enhancement or an increasing energy deficit, while ${\cal U}<0$
indicates that the inhomogeneity is decreasing (note that from its
definition (\ref{eq:sudef}) ${\cal U}$ is a tangent in the original
phase space $\{X,Y\}$, thus $-\infty<{\cal U}<\infty$). One is
therefore interested in the nature of the critical points (if any) on
the $\Omega=0$ or the $w=0$ axis (the final state of the cosmological
dynamical system): a stable node on the ${\cal U}>0$ semi-axis will
indicate that the given perturbation mode will indefinitely grow, thus
giving a Laplace instability; a stable node on the ${\cal U}<0$
semi-axis will indicate that the given perturbation mode will finally
decay, i.e. the system is stable with respect to that perturbation; a
stable node on ${\cal U}=0$ means that the perturbation will
asymptotically approach a constant value, i.e. the system is Liapunov
unstable, but still stable in the sense of Laplace (structural
instability \cite{bi:AP}). Finally, if there are no critical points on
the $\Omega=0$ (or the $w=0$) axis, the perturbation maintains a sound
wave character at any time.

A discussion of the system (\ref{eq:suev}) (\ref{eq:3}) can be
found in \cite{bi:wos} (and reference therein);
a discussion of
the flaws of the application of the analysis given in \cite{bi:wos}
to the perturbation equations of Bardeen \cite{bi:bardeen} and Ellis -
Bruni
\cite{bi:EB}, \cite{bi:EBH}, \cite{bi:BDE}
(see also \cite{bi:WK}) is given in \cite{bi:bruni}.

\section{Dynamics of the dust\hs radiation models}\label{sec:dyndr}

We shall now apply the general method outlined in the previous section
to the case of uncouple dust and radiation. In the following, we shall
normalize the scale factor at equidensity of dust and radiation,
introducing $S=a/a_E$. An analysis of the phase--space of FLRW models
containing dust and radiation has been recently given in \cite{bi:ER}
(see also \cite{bi:harrison1} and \cite{bi:padma}); here we simply
review well known results, that are needed for the perturbation
analysis, with emphasis on a useful parameterization.

\subsection{The background}\label{sec:back}

Since the two fluids are uncoupled, we have separate energy
conservation, with $\mu_d=\sfrac12\mu_E S^{-3}$ and
$\mu_r=\sfrac12\mu_E S^{-4}$, where $\mu_E$ is the {\it total} energy
density at equidensity. Then the total energy density $\mu=\sfrac12
\mu_E(S^{-3} +S^{-4})$ is also conserved, while the total pressure is
that of the radiation component, $p=p_r=\sfrac13 \mu_r=\sfrac16 \mu_E
S^{-4}$; therefore

\be w=\frac{1}{3(S+1)}\;. \label{eq:w1} \ee For two non interacting
fluids the quantity $c_s^2=\frac{\dot{p}}{\dot{\mu}}$ is only formally
the speed of sound: for dust and radiation we have \be
c_s^2=\frac{4}{3(4+3S)}\;. \label{eq:cs} \ee At equidensity $S=1$,
$w=1/6$, and $c_s^2=4/21$. Clearly, (\ref{eq:w1}) can be inverted to
give \be S=\frac{1-3w}{3w}\;, \label{eq:S} \ee and then the expansion
of the universe model we consider is parameterized by $w$, with $w$
($1/3\geq w\geq 0$) varying from a pure radiation dominated
($t\rightarrow 0$) to a pure matter dominated ($t\rightarrow \infty$)
phase.

Note that (\ref{eq:w1}), with (\ref{eq:cs}), is in practice an integral
for (\ref{eq:w}); also, given (\ref{eq:w1}) we can integrate
(\ref{eq:2}):
\be
\Omega= \frac{\Omega_E (S+1)}{\Omega_E(S+1) +2(1-\Omega_E)S^2}\;;
\label{eq:omega}
\ee since today $S_0=a_0/a_E=1+z_E\gg 1$, the density parameter at
equidensity is $\Omega_E\simeq \left(1
+\frac{(1-\Omega_0)}{2S_0\Omega_0}\right)^{-1}$. Clearly, equation
(\ref{eq:omega}) can also be inverted to parameterize the expanding
model with $\Omega$. However in the following we shall find convenient
the parameterization in $w$, which will allow us to give a unified
treatment of flat and open models: from (\ref{eq:S}) and (\ref{eq:cs})
we have \be c_s^2=\frac{4w}{3(1+w)}\;, \label{eq:cs1} \ee and this
with (\ref{eq:w}) give \be w' = w \left( w- \sfrac13 \right)\;,
\label{eq:w2} \ee where hereafter the prime stands for the derivative
with respect to $\tau=ln S^3$.

 From (\ref{eq:w2}) and (\ref{eq:3}) we get \be \Omega=
\frac{3\Omega_E w}{3\Omega_Ew +2(1-\Omega_E)(1-3w)^2}\;,\label{eq:ow}
\ee which can be used in expressions for $\xi$ and $\psi$ to obtain
these coefficients as functions of $w$ only, thus reducing the
effective phase space needed for the stability analysis to
$\{\su,w\}$, i.e.  that of the plane autonomous system
(\ref{eq:suev}), (\ref{eq:w2}). We note also that the system
(\ref{eq:w2}), (\ref{eq:3}), that describes the evolution of the
background universe model is a plane autonomous subsystem in the full
space $\{\su,\Omega,w\}$: this will always be a property of a
dynamical system describing a perturbed universe, since the basic
assumption of the perturbation analysis is that of neglecting the
backreaction of the perturbations on the dynamics of the background
model.

Finally, for completeness, we note that also the total energy density
is given by \be \mu=\frac{27}{2}\mu_E\frac{w^3}{(1-3w)^4}\;.
\label{eq:mu} \ee This obviously also follows from the conservation
equation $\mu'=-\mu(1+w)$, integrated using (\ref{eq:w2}); and
$\mu\rightarrow \infty$ as $w\rightarrow 1/3$, $\mu\rightarrow 0$ as
$w\rightarrow 0$.

\subsection{The perturbed model}\label{sec:pert}

We shall now consider the dynamics of the perturbed dust\hs radiation
models.

First, we have to specify a perturbation measure $X$, i.e. the
variable appearing in (\ref{eq:1}) and (\ref{eq:X}). Here, we shall
focus on density perturbations, and we shall take $\Delta=a^2(\n^2\mu
)/\mu$ as our fundamental variable,
as was originally defined in \cite{bi:EBH} (see also
\cite{bi:WK}) following the covariant
approach to perturbation
introduced in \cite{bi:EB}.

This covariant quantity is an exact measure of density inhomogeneity,
it is scalar and  locally defined. With respect to a FLRW background
$\Delta$ is a gauge\hs invariant variable that, once expanded at first
order in the perturbations, is proportional to the Bardeen variable
$\varepsilon_m$ (see \cite{bi:BDE});
thus its components with respect
to an orthonormal set of scalar harmonic functions are proportional to
the density perturbation in the comoving gauge (see \cite{bi:bardeen};
for a comprehensive
treatment of perturbations in this gauge see \cite{bi:LL}).

In the following, we shall restrict our attention to the harmonic
components of $\Delta$, where the scalar harmonics $Q$ are defined by

\be \n^2 Q= -\frac{k^2}{a^2} Q\;,\label{eq:helmotz} \ee $\n^2$ is the
Laplace operator in the 3-surface of constant curvature and $k\geq 0 $
for $K=0$, but $k\geq 1$ for $K=-1$ \cite{bi:harrison}, \cite{bi:LK}.
Thus in the flat case the wavenumber $k$ is simply related to the
physical scale of the perturbation i.e. its wavelength $\lambda=2\pi
a/k$, but this is not the case for the open models. However $k$ can
always be taken as invariantly characterizing the scale of the
perturbation, and we shall do so.

Since the harmonic component $\Delta^{(k)}$  of $\Delta$  and
$\varepsilon_m$ are just proportional \cite{bi:BDE}, in the following
the perturbation measure $X$ appearing in (\ref{eq:1}) and (\ref{eq:X})
can be identified either with $\Delta^{(k)}$ or with $\varepsilon_m$.

In the case of a mixture of dust and radiation, in general the
evolution equation for $X$ is coupled to the evolution equation for an
entropy perturbation variable (see \cite{bi:KS};
\cite{bi:DBE}; \cite{bi:LS} and also
\cite{bi:padma}). However the coupling is important
only at small
scales: here we shall only focus on perturbations at scales of the
order of  or larger than the Hubble radius $H^{-1}_E$ at equidensity, thus
we shall restrict to purely adiabatic modes, i.e.  solutions of
(\ref{eq:1}), neglecting the coupling with the entropy perturbation.

In general the coefficients $\alpha$ and $\beta$ in (\ref{eq:1}) are given
 by (see \cite{bi:EBH}; \cite{bi:BDE}; \cite{bi:padma})

\begin{eqnarray}
  \alpha & = & (2 +3c_s^2 -6w) H\;, \\ \beta & = &
  -\left[\left(\sfrac12 +4w -\sfrac32 w^2 -3c_s^2\right)\kappa \mu +
  12 (c_s^2 -w) \frac{K}{a^2}\right] +c_s^2
  \frac{k^2}{a^2}\;;\label{eq:beta}
\end{eqnarray} using (\ref{eq:cs1}) for the dust\hs radiation
background we get

\begin{eqnarray}
  \alpha & = & 2 \frac{(1-3w^2)}{(1+w)} H \;, \\ \beta & = &
  -\left[\frac{3}{2}\left( 1 +\frac{w^2(5-3w)}{(1+w)}\right)\Omega H^2
  +4 \frac{w(1-3w)}{(1+w)}\frac{K}{a^2}\right] +\frac{4w}{3(1+w)}
  \frac{k^2}{a^2}\;.\label{eq:beta1}
\end{eqnarray}  From this, we obtain
\begin{eqnarray}
  \psi(w,\Omega) & = & \frac{1}{6}(1+3w)(1-\Omega)
  +\frac{1-6w-15w^2}{6(1+w)}\;, \label{eq:psi} \\ \xi(w,\Omega) & = &
  -\frac{1}{6}\left( 1 +\frac{w^2(5-3w)}{1+w}\right)\Omega
  +\frac{4w(1-3w)}{9(1+w)}(1- \Omega)+\Xi_K (w,k)\;, \label{eq:xi}
\end{eqnarray} where $\Xi_K
(w,k)=\frac{4w}{27(1+w)}\frac{k^2}{a^2H^2}$ is a function that takes a
different form, as function of $w$, depending on the curvature $K$: in
an open universe $a^2H^2=(1-\Omega)^{-1}$, but in the flat case we
cannot use the Friedmann equation to substitute for $a^2H^2$. Rather,
we use it to substitute for $3H^2=\kappa\mu$, and $\mu$ is given by
(\ref{eq:mu}), while here $a=Sa_E$ and $S$ is given by (\ref{eq:S}).
Thus \be \label{eq:xi0}
\begin{array}{lll}
  K=0 & \Xi=\frac{(1-3w)^2}{1+w}\Xi_0 k^2\;, & \Xi_0=\frac{8}{81a^2_E
    H_E^2}\;, \\ K=-1 & \Xi=\frac{4w}{27(1+w)} (1-\Omega) k^2\;, &
  \mbox{}
\end{array} \ee where $3H_E^2=\kappa\mu_E$, and for open models we can
substitute for $\Omega=\Omega(w)$ from (\ref{eq:ow}) in all the
previous expressions.  It is clear from (\ref{eq:xi0}) that in the
space $\{\su,w,\Omega\}$ the function $\Xi_{K}$ is not continuous (for
$k\not=0$) on the plane $\Omega=1$, except on the line $w=1/3$,
$\Omega=1$; this discontinuity of $\Xi_{K}$ will play an important
role in the behaviour of perturbations.

\section{Results}\label{sec:res}

We shall now summarize the results we obtained from the analysis of
the dynamical system given by (\ref{eq:suev}), (\ref{eq:w2}) and
(\ref{eq:3}), with coefficients $\psi$ and $\xi$ given by
(\ref{eq:psi}) and (\ref{eq:xi}), first restricting to the flat models
and then considering the open ones.

\subsection{The flat models}\label{sec:flat}

The flat perturbed models are described by the subsystem
(\ref{eq:suev}), (\ref{eq:w2}) substituting $\Omega=1$ in $\psi$
(\ref{eq:psi}) and $\xi$ (\ref{eq:xi}) and using $\Xi$ for $K=0$ given
in (\ref{eq:xi0}).   From this latter we see that for flat models
there is a characteristic scale that, as we shall see, is related to
the late time behaviour of the perturbations: this is the Hubble
radius at equidensity $H^{-1}_E$. It is therefore convenient to define
\be k_E=\frac{k}{a_E H_E}
=2\pi\frac{H_E^{-1}}{\lambda_E}\;,\label{eq:ke} \ee which represents a
wavenumber normalized at equidensity:
$k_E=2\pi\Leftrightarrow\lambda_E=H^{-1}_E$ i.e. $k_E=2\pi$
corresponds to a perturbation wavelength that enters the horizon at
equivalence epoch.

Stationary points eventually exist on the $w=1/3$ and $w=0$ axis, with
\be
\su_{\pm}  =  \sfrac12 (-\psi_w \pm \sqrt{\psi^2_w-4\xi_w})\;,
\label{eq:u+}
\ee where $\psi_w$, $\xi_w$ (with $w=0,1/3$) correspond to the
stationary values of $\psi$ and $\xi$.

The stationary points for the system (\ref{eq:suev}), (\ref{eq:w2}) are
given in Table \ref{tab:1}.

As we have explained at the end of section \ref{sec:dyn}, the
condition for having growing perturbations is given by the appearance
of a stable node on the positive side of the $\su$ axis.   From
(\ref{eq:u+}) we see that we need $\xi_0\leq0$ in order to have
$\su_{+}\geq 0$, i.e.  \be k_E\leq k_{EC}\;,
{}~~~~~k_{EC}=\frac{3\sqrt{3}}{4}\;.\label{eq:kc} \ee
Moreover, stationary points only exist for $\psi^2_w-4\xi_w\geq 0$, which
\begin{table}
  \vspace{0.5truecm}
\begin{tabular}{|c|c|} \hline
  \hline {\sc Point I: Unstable Node} & {\sc Point II: Saddle} \\ $
  w=1/3$,~~ $\su_{-}=-\frac{1}{3}$ & $w=1/3$,~~ $\su_{+}=\frac{2}{3}$
  \\ \hline $\lambda_{\su}=1$ & $\lambda_{\su}=-1$ \\
  $\lambda_{w}=\frac{1}{3}$ & $\lambda_w=\frac{1}{3}$\\ \\ \hline {\sc
    Point III: Saddle} & {\sc Point IV: Stable Node} \\ $w=0$,~~
  $\su_{-}=-\frac{1}{12}-\lambda_{\su}$ & $w=0$,~~
  $\su_{+}=-\frac{1}{12} -\lambda_{\su}$ \\ \hline $
  \lambda_{\su}=\sqrt{ \frac{2}{3} \left( \frac{25}{24}
    -\frac{k_E^2}{k_{EC}^2} \right) }$ & $\lambda_{\su}=-\sqrt{
    \frac{2}{3} \left( \frac{25}{24} -\frac{k_E^2}{k_{EC}^2} \right)
    }$ \\ $\lambda_w=-\frac{1}{3}$ & $\lambda_w=-\frac{1}{3}$ \\
  \hline \hline
\end{tabular}
\caption{Flat models: the four stationary points with the corresponding
  eigenvalues along the axis $\su$ and $w$ and their nature; Point III
  and IV exist only for $k_{E} \leq \frac{5}{2\protect\sqrt{6}} k_{EC}
  $.}
\label{tab:1}
\end{table}
is always satisfied only for $w=1/3$. For $w=0$ we have
$\psi_0^2-4\xi_0=\sfrac23 (\sfrac{25}{24}-k_E^2/k^2_{EC})$, i.e.
stationary points exist on this axis for perturbations of wavelength
at equivalence
$\lambda_E\geq\case{2\sqrt{6}}/{5}\lambda_{EC}<\lambda_{EC}$.  We see
from Table \ref{tab:1} that this is the same condition for the reality
of the eigenvalues, a result that follows from the fact that $w'$ does
not depend on $\su$. From the values of these eigenvalues we have
that: Point I is an unstable node, Point II is a saddle, Point III is
a saddle, Point IV is a stable node.
\begin{figure}
  \vspace{6 cm}
\caption{Flat models: phase space for the evolution of large scales
  $\lambda_E>\lambda_{EC}$ that grow unbounded ($\su_{+}>0$).}
\label{fig:1a}
\end{figure}
\begin{figure}
  \vspace{6 cm}
\caption{Flat models: phase space for the evolution of the damped
waves on small scales 
($\su_{\pm}\in \Im$).}
\label{fig:1b}
\end{figure}
\begin{figure}
  \vspace{6 cm}
\caption{Flat models: phase space for the overdamped modes ($\su_{+}<0$)
  on intermediate scales
  $\frac{2\protect\sqrt{6}}{5}\lambda_{EC}<\lambda_E<\lambda_{EC}$.}
\label{fig:1c}
\end{figure}

When for this latter we have $\su_{+}>0$, i.e. for $k_{E}$ smaller
than the {\it critical} wavenumber $k_{EC}$, generic perturbations
with physical wavelength $\lambda_{E}>\lambda_{EC}$ {\it grow
  unbounded}, with the critical perturbation scale $\lambda_{EC}$
corresponding to $k_{EC}$ in (\ref{eq:kc}), given by $\lambda_{EC}=
\frac{8\pi}{3\sqrt{3}}H_{E}^{-1}$, i.e.  $\lambda_{EC}>H_{E}^{-1}$:
this situation is illustrated in Fig. \ref{fig:1a}. Note in this
figure the special {\it saddle trajectories}: the one ending in
$\su_{-}$ represents the purely decaying mode, while the other (ending
in $\su_{+}$ as the generic trajectory) represents the purely growing
mode (the attractor); the generic trajectory represents a linear
combination of the two modes solutions of (\ref{eq:X}). Also, we
remind that the variable $\su$ is a tangent in the original phase
space $\{ X,X'\}$, so that the trajectories in the figures that start
from Point I and go to the left of the saddle trajectory exit the
figure on the left boundary, and re-enter it from the right ending in
Point IV.

Clearly, when Point IV does not exist, i.e. for
  perturbations with wavelength at equivalence $\lambda_E
<\sfrac{2\sqrt{6}}{5}\lambda_{EC}$, we are considering perturbations
that always {\it oscillate} as ``sound
waves'' (see Fig. \ref{fig:1b}); we have checked through a direct
stability analysis of the system (\ref{eq:os}) below (see section
\ref{sec:mc}) that
the amplitude of these modes decay. Again, trajectories that exit
from the left re-enter from the right: for $\lambda_E
<\sfrac{2\sqrt{6}}{5}\lambda_{EC}$  however this
happens many times, and indeed it is this that characterizes the
oscillatory behaviour of these modes in the $\{\su,w\}$ plane.

Also, for scales $k_{EC}<k_E<\sfrac{5}{2\sqrt{6}}k_{EC}$
we have $\su_{+}<0$ for Point IV, i.e. the stable node  is located
on the negative side of the
$\su$ axis: thus perturbation scales in this range
are overdamped:they decay without oscillating, as it
appears from Fig. \ref{fig:1c}. Again, the saddle trajectories
represent the two modes for (\ref{eq:X}), and the trajectories
going out from the left of Fig. \ref{fig:1c} re-enter from the right,
ending in Point IV.

Finally, we point out that for $k_E=\sfrac{5}{2\sqrt{6}}k_{EC}$ the
two stationary points on the $\su$ axis (the saddle and the stable
node in the second line of Table \ref{tab:1}) coincide, i.e. this is a
{\it fold bifurcation point} (see e.g. \cite{bi:AP}). The corresponding scale
$\lambda_E=\sfrac{2\sqrt{6}}{5}\lambda_{EC}=\sfrac{16\sqrt{2}}{15}\pi
H^{-1}_E \simeq 4.7 H^{-1}_E$ is quite larger than the Hubble radius
at equidensity; thus we can consider scales
$\lambda_E<\sfrac{2\sqrt{6}}{5}\lambda_{EC}$, as in Fig. \ref{fig:1b},
such that still $\lambda_E>H^{-1}_E$: this partially justifies
our assumption of purely adiabatic perturbations (cf. \cite{bi:MFB}).

Thus we have three different evolution regimes for the adiabatic perturbation
modes of a mixture of uncouple dust and radiation in a flat universe,
depending on their wavelength:
\begin{enumerate}
\item large scale perturbations that grow unbounded, giving
  instability
\item intermediate scale perturbations that are overdamped, i.e.
  decaying without oscillating
\item small scale damped perturbations which oscillate like sound
  waves while their amplitude decays,
\end{enumerate}
\begin{table}
  \vspace{0.5truecm}
\begin{tabular}{|c|c|c|}\hline
  \hline {\sc Instability} & \multicolumn{2}{c|} {\sc Stability} \\
  \hline $\lambda >\lambda_{EC}$ &
  $\sfrac{2\sqrt{6}}{5}\lambda_{EC}<\lambda_E<\lambda_{EC}$ &
  $\lambda_E < \sfrac{2\sqrt{6}}{5} \lambda_{EC}$ \\ grow unbounded &
  critical damping & damped waves \\ \hline \hline
\end{tabular}
\caption{\label{tab:2}
  Flat models: summary of the three different evolution regimes for
  different perturbation wavelengths.}
\end{table}
as summarized in Table \ref{tab:2}.

\subsection{The open models}\label{sec:op}

The open models are in principle described by the full 3-dimensional
system given by (\ref{eq:suev}), (\ref{eq:w2}) and (\ref{eq:3}), with
trajectories in the $\{\su,w,\Omega\}$ space. However, given
$\Omega=\Omega(w,\Omega_E)$ (\ref{eq:ow}), we can substitute for
$\Omega$ in the expressions for $\psi$ (\ref{eq:psi}), $\xi$
(\ref{eq:xi}) and $\Xi$ for $K=-1$ (\ref{eq:xi0}), and restrict our
analysis to the 2-dimensional system (\ref{eq:suev}), (\ref{eq:w2}).
In doing this, we are selecting one particular open model in the class
parameterized by $\Omega_E$: there is no loss of generality in doing
this, as the dynamical properties of the models in this class (as
specified by the stationary points in their phase space) do not depend
on $\Omega_E$, i.e. for a given wavenumber $k$ the phase space
evolution of all the models in the class is qualitatively the same.
 From the point of view of the geometry of the phase space
$\{\su,w,\Omega\}$ we are looking at the trajectories in the
2-dimensional surface specified by $\Omega_E$ in this space: we shall
then consider the projection of this surface with its trajectories in
the $\{\su,w\}$ plane, as
depicted in Fig. \ref{fig:2}.
\begin{figure}
  \vspace{8 cm}
\caption{Open models: the phase space evolution is independent of the
  wavelength, and all perturbations tend to a constant value
  ($\su_{+}=0$).}
\label{fig:2}
\end{figure}

It is clear from this picture and Table \ref{tab:3} that, contrary to
what we have seen for the flat models, the existence, position and the
nature of Points I--IV do not depend on the wavenumber $k$, so that
all open models share the same dynamical history. This depends on the
vanishing of the function $\Xi$ in the limit $w\rightarrow 0$, which
also implies $\Omega\rightarrow 0$ (because we are moving on the
surface specified by $\Omega_E$). Note however that, if we do not
substitute for $\Omega$ from (\ref{eq:ow}) in the 3-dimensional
system, and we take the limit $\Omega\rightarrow 0$, this does not
give the 2-dimensional system for flat models: this fact is due to the
discontinuity (remarked at the end of section \ref{sec:flat}) of the
function $\Xi$ on the surface $\Omega=1$, where $\Xi$ for $K=0$ do not
vanish in general, except for $w=1/3$ or $k=0$.  In some way, this
fact can be seen as an example of "fragility" in cosmology
\cite{bi:reza}.

 From table \ref{tab:2} and Fig. \ref{fig:2} we see that Point IV is a
stable node located at $\su=0$.
The generic trajectory ends up in this point, either directly from
Point I, or first going out from the left boundary of the picture and
then re-entering from the right. The saddle trajectory starting form
Point I and ending in Point III represents the evolution of the purely
decaying mode, and the saddle trajectory starting from Point III and
ending in Point IV represents the purely growing
mode\footnote{The terminology {\it growing} and {\it decaying} is
purely conventional: it is adopted here because it is standard in the
literature to refer in this way to the two modes, as they are
effectively growing and decaying e.g. in flat pure dust models.
We stress again that the "growing" mode is actually growing only when
$\su_{+}>0$.} (attractor).
The fact that
Point IV is located in $\su=0$ means that all the perturbation modes
evolve up to a constant value, and then freeze--in. We can say that open
models are Liapunov unstable (the generic perturbation modes do not
decay), but Laplace stable (the perturbations do not grow unbounded), a
situation that can be described as {\it structural instability} of the
dynamical system \cite{bi:AP}.

A direct analysis of
the  system (\ref{eq:os}) below (section \ref{sec:mc}) shows  indeed that
the appearance of a fixed point on $\su=0$ for the system
(\ref{eq:suev}), (\ref{eq:w2}) corresponds to the vanishing of one of
the eigenvalue for the corresponding fixed point in the phase space
$\{X,X',w\}$. While this happens only for $k_E=k_{EC}$ in flat models,
it is a generic characteristic for any $k$ in open universes. A comparison
of Fig. \ref{fig:2} and Fig. 1a in \cite{bi:bruni} shows that open
dust-radiation models and open pure dust models share the same
stability properties. It appears that the curvature of the background
dominates also the evolution of the perturbations: for
dust radiation models there is a conspiracy between curvature and
pressure, such that curvature has an opposite effect on small and
large scale: with respect to the behaviour in flat models, it avoids the
 oscillation of small scales perturbations
($\lambda_E<\sfrac{2\sqrt{6}}{5}\lambda_{EC}$) and damps the growth of
those on large scales($\lambda_E>\sfrac{2\sqrt{6}}{5}\lambda_{EC}$).

\begin{table}
\vspace{0.5truecm}
\begin{tabular}{|c|c|} \hline
  \hline {\sc Point I: Unstable Node} & {\sc Point II: Saddle} \\ $
  w=\frac{1}{3}$,~~ $\su_{-}=-\frac{1}{3}$ & $w=\frac{1}{3}$,~~
  $\su_{+}=\frac{2}{3}$ \\ \hline $\lambda_{\su}=1$ &
  $\lambda_{\su}=-1$ \\ $\lambda_{w}=\frac{1}{3}$ &
  $\lambda_w=\frac{1}{3}$\\ \\ \hline {\sc Point III: Saddle} & {\sc
    Point IV: Stable Node} \\ $w=0$, ~~ $\su_{-}=-\frac{1}{3}$ &
  $w=0$,~~ $\su_{+}=0$ \\ \hline $ \lambda_{\su}=\frac{1}{3} $ &
  $\lambda_{\su}=- \frac{1}{3} $ \\ $\lambda_w=-\frac{1}{3}$ &
  $\lambda_w=-\frac{1}{3}$ \\ \hline \hline
\end{tabular}
\caption{\label{tab:3}
  Open models: the four stationary points with the corresponding
  eigenvalues along the axis $\su$ and $w$ and their nature.}
\end{table}

\subsection{Comparison with Jeans instability}

We can apply the Jeans instability criterion directly to equation
(\ref{eq:1}): this implies that gravitational collapse of a given
perturbation mode will occur if $\beta<0$, i.e. if $k<k_J$, where for
the dust--radiation models $\beta$ is given by (\ref{eq:beta1}), and
the corresponding $k_J$ is \be \frac{{k}^{2}_{J}}{a^2} =
\frac{3(1+w)}{4w} \left[\frac{3}{2}\left(1+
\frac{w^2(5-3w)}{1+w}\right)\Omega H^2
+\frac{4w(1-3w)}{1+w}\frac{K}{a^2}\right]\;.\label{eq:kj} \ee At
equidensity ($w=1/6$) this gives \be
\frac{k^2_{JE}}{a_E^2}=\frac{279}{32}\Omega_E H_E^2
+\frac{3}{2}(\Omega_E-1)H^2_E\;; \ee in the real universe
$\Omega_E\approx 1$ and the contribution to $k_{JE}$ from the
curvature term (the last in the equation above) is completely
negligible. Thus in a flat universe ($K=0$) we have \be
\lambda_{JE}\equiv \frac{2\pi
  a_E}{k_{JE}}=\frac{8\pi}{3}\sqrt{\frac{2}{31}} H^{-1}_E\; \ee for
the Jeans scale at equidensity. Then a comparison with the critical
scale $\lambda_{EC}$ defined in section \ref{sec:flat} gives \be
\lambda_{JE}=\sqrt{\frac{6}{31}}\lambda_{CE}\;,\label{eq:ratio} \ee
i.e. for flat models the stability criteria used in section
\ref{sec:flat} give a critical scale $\lambda_{EC}$ for instability
which is larger than the corresponding Jeans scale at equidensity by a
factor of $2$, i.e. $\lambda_{EC}\approx 2.3 \,\lambda_{JE}$. The fact
that the values of these two scales are relatively close appears
physically meaningful, and in a way obvious, since both scales are
somehow defined through the same differential equation.  However, a
comparison of (\ref{eq:ratio}) with the analysis of section
\ref{sec:flat} shows that there are perturbations with
$\lambda_E>\lambda_{JE}$ that decay: those with wavelength
$\lambda_{JE}<\lambda_E<\sfrac{2\sqrt{6}}{5}\lambda_{EC}$ are damped
oscillation, while those with
$\lambda_{JE}>\lambda_E>\sfrac{2\sqrt{6}}{5}\lambda_{EC}$ are
overdamped. It is immediate to show that today the critical scale
corresponding to $k_{EC}$ is \be
\lambda_{0C}=\frac{4\pi}{3}\sqrt{\frac{2}{3}(1+z_E)^{-1}} H_0^{-1}\;,
\ee i.e. about $67.3 h^{-1} Mpc$ in a flat universe with $1+z_E\approx
4.3 10^{-5} h^{-2}$ (see e.g. \cite{bi:KT}). Even if this value may be
an artifact of our simplifying assumptions, e.g.  the fact that we
have neglected isocurvature modes at all times and we have treated
radiation as a perfect fluid even at small scales, we believe that the
discrepancy between our critical scale and the Jeans scale in the same
universe model is a general feature that deserves further
investigation in order to consider possible effects for models of
structure formation in the universe.

Finally, it is interesting to consider the limit  $w\rightarrow 0$
of $k_J$. For flat models, one gets from (\ref{eq:kj}) that in this
limit $k_{J}=\frac{3\sqrt{3}}{4}a_E H_E$, i.e. we recover $k_{EC}$
(\ref{eq:kc}) in this limit. However the same limit for open models
gives a value for $k_J$ that: {\it a)} is real only for $\Omega_E
>16/25$, a result that, although satisfied in the real universe,
appears spurious for the theory; {\it b)} gives the false impression
that there could be growing and oscillating modes also for open
models, contrary to what we have shown in the previous section.

\subsection{Metric and curvature perturbations} \label{sec:mc}

In the previous sections we have given the results of the analysis of
the dynamical system $\{\su,w\}$ (\ref{eq:suev}), (\ref{eq:w2}) for
flat and open models, and inferred conclusions on the evolution of
density perturbations, represented by the harmonic component $X$ (see
section \ref{sec:pert}). As we have already pointed out, there is a
particular relation between the location and character of the critical
points in the phase space $\{\su,w\}$, and the character of the
corresponding point in the original phase space
$\{X,X',w\}$\footnote{Here we assume that, for open models, we are
  using the function $\Omega=\Omega(w,\Omega)$ (\ref{eq:ow}), so that
  the further dimension $\Omega$ in the phase space is suppressed.}.
Indeed, it is immediate that the original system ($Y\equiv X'$)
\be\label{eq:os} \left\{
\begin{array}{rcl}
  X' & = & Y \\ Y' & = & -\psi Y -\xi X \\ w' & = &
  w\left(w-\sfrac13\right)
\end{array} \right.  \ee admits only two fixed points for $\xi\not=0$:
Point ${\rm A}\equiv\{X=0,Y=0,w=1/3\}$ and Point ${\rm
  B}\equiv\{X=0,Y=0,w=0\}$. Then it is easy to see that the
eigenvalues at these points along the principal directions in the
$\{X,Y\}$ planes equate the roots $\su_{\pm}$ of (\ref{eq:suev}), i.e.
$\lambda_{\pm}=\su_{\pm}$.  Point A is the same for flat and open
models ($\lambda_{-}=-1/3$, $\lambda_{+}=2/3$, $\lambda_w=1/3$), while
(\ref{eq:xi}), (\ref{eq:xi0}) and the analysis in section \ref{sec:op}
shows that for open models $\xi=0$ for $w=0$, and thus Point B
degenerates into a line (the $Y=0$ axis): $\lambda_{+}=\su_{+}=0$ in
this case, and we have {\it structural instability} \cite{bi:AP}. For
flat models, the same happens for $k_E=k_{EC}$, as already pointed
out.

Having clarified
the relation between the roots $\su_{\pm}$ of the system
(\ref{eq:suev}), (\ref{eq:w2}) and the eigenvalues at the fixed points
of (\ref{eq:os}), we can now turn to the asymptotic evolution of $X$.
It is clear from the definition of $\su=Y/X$
 that around the roots $\su_{\pm}$ the evolution of $X$ is given by
\be
X'=\su_{\pm} X\;, ~~~\Rightarrow ~~~X_{\pm}\sim S^{3\su_{\pm}}\;.
\label{eq:xev}
\ee When $\su_{+}=0$ we have $X_{+}=const.$ (cf. Table \ref{tab:3});
for $w\rightarrow 0$ we recover the well known constant mode for
matter dominated open models: here the same mode is found for the
critical scale $k_E=k_{EC}$ in flat models.

In tables \ref{tab:4} and \ref{tab:5} we give the asymptotic
behaviour for $X$, for  a metric perturbation $\Phi_N$ and a
dimensionless curvature perturbation scalar $E/\Theta^2$. In
particular, in Table \ref{tab:4} we consider flat models in the limit
$w\rightarrow 0$, giving the asymptotic behaviour of various variables
as functions of the scale factor $S$ and in order of increasing
wavelength $\lambda_E$. In Table \ref{tab:5} we give the asymptotic
solutions of open models around Points I--IV: as for $w\rightarrow
1/3$ the universe is radiation dominated and also $\Omega\rightarrow
1$, then  $S\sim t^{\frac{1}{2}}$ in this limit, and the asymptotic
solutions around Points I and II are in common with flat models. In
the limit $w\rightarrow 0$ the universe models are matter dominated,
and then in flat models $S\sim t^{\frac{2}{3}}$, while in open models
$S\sim t$.

\begin{table*}
  \vspace{0.5truecm}
\begin{tabular}{|c|c|c|c|c|c|} \hline
  \hline {\sc Quantity} &
  {$\lambda_E<\frac{5}{2\sqrt{6}}\lambda_{EC}$} &
  {$\lambda_E=\frac{5}{2\sqrt{6}}\lambda_{EC}$} &
  {$\lambda_{EC}>\lambda_E>\frac{5}{2\sqrt{6}}\lambda_{EC}$} &
  {$\lambda_E=\lambda_{EC}$} & {$\lambda_E\gg\lambda_{EC}$ } \\
  \rule{0cm}{1.5truemm} \\
\hline
$X_{\pm} $ & $S^{-\frac{1}{4}\pm \imath Q} $ &
 $S^{-\frac{1}{4}} $ & $S^{-\frac{1}{4}\pm P}  $ &
 $S^{-\frac{1}{2}}\;, ~const. $ & $ S^{-\frac{3}{2}}\;, ~S $ \\
\hline
 $\Phi_{N\pm} $ & $S^{-\frac{5}{4}\pm \imath Q} $ &
 $S^{-\frac{5}{4}} $ & $S^{-\frac{5}{4}\pm P} $ &
$S^{-\frac{3}{2}},~S^{-1} $ & $S^{-\frac{5}{2}},~const. $ \\
\hline
 $E_{\pm}/\Theta^2 $ & $S^{-\frac{1}{4}\pm \imath Q} $ &
 $S^{-\frac{1}{4}} $ & $S^{-\frac{1}{4}\pm P}  $ &
 $S^{-\frac{1}{2}}\;, ~const. $ & $ S^{-\frac{3}{2}}\;, ~S $ \\
\hline
\hline
\end{tabular}
\caption{\label{tab:4}
  Flat models: asymptotic behaviour for $w\rightarrow 0$ (i.e. around
  Points III--IV) of $X$, $\Phi_N$ and $E/\Theta^2$, in order of
  increasing wavelengths, as function of the scale factor $S$ ($S\sim
  t^{\frac{2}{3}}$ for $w\rightarrow 0$).  The decaying ($-$, Point
  III) and growing ($+$, Point IV) modes are distinguished either by
  the $\pm$ or presented in order: they coincide for the critical
  damping scale $\lambda_E=\frac{5}{2\protect\sqrt{6}}\lambda_{EC}$;
  $Q=\left[\frac{2}{3}\left(k_E^2/k^2_{EC}-
  \frac{25}{24}\right)\right]^{\frac{1}{2}} $, and
  $P=\left[\frac{2}{3}\left(\frac{25}{24}
  -k_E^2/k^2_{EC}\right)\right]^{\frac{1}{2}}$.}
\end{table*}

\begin{table}
  \vspace{0.5truecm}
\begin{tabular}{|c|c|c|c|c|} \hline
  \hline {\sc Quantity} & {\sc Point I} & {\sc Point II} & {\sc Point
    III} & {\sc Point IV} \\ \rule{0cm}{1.5truemm} \\ \hline $X $ &
  $S^{-1} $ & $S^{2} $ & $S^{-1} $ & $const. $ \\ \hline $\Phi_N$ &
  $S^{-3} $ & $const. $ & $S^{-2} $ & $S^{-1} $ \\ \hline
  $\frac{E}{\Theta^2} $ & $S^{-1} $ & $S^{2} $ & $S^{-2} $ & $S^{-1} $
  \\ \hline \hline
\end{tabular}
\caption{\label{tab:5}
  Open models: asymptotic behaviour around Points I--IV for $X$,
  $\Phi_N$ and $E/\Theta^2$, as function of the scale factor $S$. For
  $w\rightarrow 1/3$ $S\sim t^{\frac{1}{2}}$, and Point I and II are
  in common with flat models, while for $w\rightarrow 0$ $S\sim t$.
  Points I and III correspond to the decaying mode, and II and IV to
  the growing mode.}
\end{table}

In the following we shall outline the relation between the
gauge--invariant metric potential $\Phi_N$, the curvature variable
$E$, and the density perturbation $\Delta$ (see section
\ref{sec:pert}): more details can be found in \cite{bi:BDE} and
references therein . Let $E_{ab}$ be the electric part of the Weyl
tensor\footnote{Latin indices are 4-dimensional ($0,1,2,3$), and greek
  indices 3-dimensional ($1,2,3$).} $C_{acbd}$: $E_{ab}\equiv
C_{acbd}u^cu^d$, where $u^a$ is the 4-velocity of matter; then to
first order \be a^2 \n^b\n^aE_{ab}=\frac{\kappa\mu}{3}\Delta\;,
\label{eq:const} \ee where $\n_a$ is a covariant derivative orthogonal
to $u^a$. The analogue of $E_{ab}$ in Newtonian theory is the tidal
field $E_{\alpha\beta} =\nabla_{\alpha\beta} \phi$, where $\phi$ is
the Newtonian potential and $\nabla_{\alpha\beta}\phi\equiv
\na^2\phi-\sfrac13 \delta_{\alpha\beta}\phi$. Then it is possible to
show that to linear order a similar formula holds in relativistic
perturbation theory, i.e. $E_{\alpha\beta}=\na_{\alpha\beta}\Phi_N$
for the scalar part of $E_{ab}$ (all the $\{0,0\}$ and $\{0,\alpha\}$
components are second order). Moreover the field $\Phi_N$, which play
here the role of a gauge--invariant analogue of the Newtonian
potential, is just $\Phi_N=\sfrac12 (\Phi_A-\Phi_H)$, where $\Phi_A$
and $\Phi_H$ are the gauge invariant metric perturbations defined in
\cite{bi:bardeen} ($\Phi_A=-\Phi_H$ for perfect fluids). Then using
$E_{\alpha\beta}=\na_{\alpha\beta}\Phi_N$ in (\ref{eq:const}), the
harmonic decomposition (\ref{eq:helmotz}) and
$\Delta^{(k)}=-k^2\varepsilon_m$, we get equation (4.3) of
\cite{bi:bardeen}: \be 2(3K-k^2)\Phi_N =\kappa a^2\mu
\varepsilon_m\;.\label{eq:const1} \ee Then this relation can be used
to determine the asymptotic evolution of the gauge--invariant metric
potential $\Phi_N$, and also that of amplitude
$E=\sfrac12\sqrt{E_a{}^b E_b{}^a}$ of the tidal field $E_{ab}$: indeed
from a comparison of (\ref{eq:const}) and (\ref{eq:const1}) the
harmonic components of $E$ and $\Phi_N$ are related by \be
E=\frac{1}{2} a^{-2}k^2\Phi_N\;.  \ee Then it is usual to consider a
dimensionless scalar to measure the relative dynamical significance of
a given field using $\Theta$ to take into account the expansion (see
e.g.  \cite{bi:goode}): thus in the case of $E$ we consider
$E/\Theta^2$, for which we have \be \frac{E}{\Theta^2}\sim
\frac{a^{-2}\Phi_N}{\Theta^2}\sim
\frac{\kappa\mu}{\Theta^2}\varepsilon_m \sim \Omega \varepsilon_m\;.
\ee Hence, for example for a flat model $\Phi_N\sim const.$ for very
large scales, but the relative amplitude of the tidal field
$E/\Theta^2\sim \Delta\sim a$ grows unbounded (cf. \cite{bi:PKSD}).

\section{Conclusions}\label{sec:conc}

In this paper we have considered the stability properties of FLRW
models with uncouple Cold Matter (dust) and radiation.  We have
considered the perturbed models as dynamical systems described by an
evolution equation for a gauge--invariant density perturbation
variable coupled with the equations governing the evolution of
relevant background variables: the pressure--energy density ratio
$w=p/\mu$ and the density parameter $\Omega$. For the subsystem
describing flat models, given by $\Omega=1$, we deal with a planar
autonomous system, and for the open models we have shown that we can
also restrict the analysis to a planar system for each particular
value of the density parameter at equidensity $\Omega_E$.

The analysis of flat and open models gives different results: flat
models admit unstable perturbation modes, while open models are
Laplace stable (but Liapunov unstable) with respect to perturbations,
irrespective of their size, ie. the perturbations freeze--in at a
constant value (structural instability).
Qualitatively, both results can be expected on the
basis of the results of a more traditional study of the behaviour of
the perturbations in simple models. The final fate of the
perturbations in open models appears to be dominated by the background
curvature, which  governs the background expansion at late times:
therefore the stability properties of the dust--radiation models are
the same as those of a pure dust model, irrespective of the size of
the perturbation and of the radiation content of the given model.

Instead, we find more interesting features for flat models: in this
case each model has  a characteristic critical invariant wavenumber
$k_{EC}$ which depends on the proportion of matter and radiation (e.g.
at present), and the corresponding scale $\lambda_{EC}$ determines the
transition from stable to unstable modes. The scale $\lambda_{EC}$ is of the
order of the Hubble radius $H^{-1}_E$ at the equidensity of matter and
radiation, $H^{-1}_E$ being the only scale entering the background
model. The present value of this transition scale is
$\lambda_{0C}=(1+z_E)\lambda_{EC}$ and depends on the
actual present proportion of matter and radiation, for which it is of the
order of $67 Mpc$.
Therefore the stability properties of dust--radiation flat models
are a mixture of the properties of pure dust and pure radiation
models, being well known that perturbations in a flat dust models grow
unbounded irrespective of their size, while all perturbation scales in a
radiation model always enter an oscillatory (sound waves) regime after
they enter the sound horizon (see e.g. \cite{bi:bardeen}).

However, we
actually find a  structure in the stability properties of flat
dust--radiation models which is more interesting than expected,
because we find
three different regimes for the evolution of the perturbations.
Perturbations on scales $\lambda_E > \lambda_{EC}$ grow unbounded
(unstable modes), while perturbations in the range
$\sfrac{2\sqrt{6}}{5}\lambda_{EC}<\lambda_E<\lambda_{EC}$ are
overdamped; finally, perturbations on scales $\lambda_E <
\sfrac{2\sqrt{6}}{5}\lambda_{EC}$ are damped, i.e. they oscillate as
``sound waves'' while decaying.

The stability properties of perturbed FLRW models that we have found
should be taken {\it cum grano salis} with respect to the problem of
structure formation in the universe, because we have assumed adiabatic
perturbations at all times, neglecting isocurvature modes that are in
principle important at small scales and late times, and we have
neglected photon diffusion, treating radiation as a perfect fluid
irrespective of the scale of the perturbation (in particular, the
oscillatory behaviour of small scales in flat models is probably an
artifact of this assumption). However, we believe
that these assumptions should not question the validity of our main
results: {\it a)} in a given universe model (including the assumptions
about the matter content and how to treat it) there are perturbations
that decay despite the fact that  their wavelength at equidensity
$\lambda_E$ is larger than the Jeans scale $\lambda_{JE}$; {\it b)}
the today size of the critical scale we have found is of the order of
$67$ Mpc, a fact that perhaps deserves further investigation regarding
its implication for structure formation in the universe.

Finally, we remark that
a  comparison of the stability properties of dust--radiation models  with the
observed small amplitude of the large scale density fluctuations
 seems to suggest that if the
spatial curvature of the universe vanishes, then we live in a special
epoch in which these perturbations  are still in a linear
regime of growth, while if we
live in a universe with negative spatial curvature the smallness of
the large scale perturbations is a characteristic of the model at all
 times and the description of the universe as an open FLRW model
is appropriate at any epoch.

Thus open models are special (at a given time) on  average, because
the density parameter $\Omega$ depends on time (this is the flatness
problem), but not from the local point of view, because large scale
structures are frozen--in, while the reverse is true for flat models,
because $\Omega=1$ at all times, but large scale perturbations grow
unbounded in these models.

\subsection*{Acknowledgements}

One of us (MB) is grateful to D.Arrowsmith and A.Polnarev for
enlightening discussions. We thank SERC (UK) and MURST (Italy) for
financial support.

\end{document}